\documentclass{svjourA}
\usepackage{blindtext}

\smartqed

\AtBeginDocument{%
  \setlength{\oddsidemargin}{\dimexpr(\paperwidth-\textwidth)/2-1in}%
  \setlength{\evensidemargin}{\oddsidemargin}%
  \setlength{\topmargin}{%
    \dimexpr(\paperheight-\textheight)/2-\headheight-\headsep-1in}%
}

\usepackage{amssymb}
\usepackage{mathptmx}       % selects Times Roman as basic font
\usepackage{helvet}         % selects Helvetica as sans-serif font
\usepackage{courier}        % selects Courier as typewriter font
\usepackage{type1cm}        % activate if the above 3 fonts are not available on your system
\usepackage{makeidx}         % allows index generation
\usepackage{graphicx}        % standard LaTeX graphics tool when including figure files
\usepackage{multicol}        % used for the two-column index
\usepackage[bottom]{footmisc}% places footnotes at page bottom

\begin{document}

\title{Quantum-gravity phenomenology \\ with primordial black holes}
\author{F. Vidotto, A. Barrau, B. Bolliet, M. Schutten and C. Weimer}
\institute{Francesca Vidotto and Marrit Schutten
\at Radboud University, Institute for Mathematics, Astrophysics and Particle Physics, P.O. Box 9010, 6500 GL Nijmegen, The Netherlands  \\ \email{F.Vidotto@science.ru.nl, M.Shutten@students.ru.nl}
\and 
Aur\'elien Barrau, Boris Bolliet and Celine Weimer 
\at Laboratoire de Physique Subatomique et de Cosmologie, Universit\'e Grenoble-Alpes, CNRS-IN2P3,
53 avenue des Martyrs, 38026 Grenoble, France \\
\email{Aurelien.Barrau@cern.ch, Boris.Bolliet@ens-lyon.fr, CelineW@kth.se}
 }
\maketitle

% Use \abstract* if it should only appear in the online version
\abstract{Quantum gravity may allow black holes to tunnel into white holes. If so, the lifetime of a black hole could be shorter than the one given by Hawking evaporation, solving the information paradox. More interestingly, this could open to a new window for quantum-gravity phenomenology, in connection with the existence of primordial black holes (PBH). We discuss in particular the power of the associated explosion and the possibility to observe an astrophysical signal in the radio and in the gamma wavelengths.
}

% Always give a unique label
% and use \cite{<label>} for cross-references
% and \cite{<label>} for bibliographic references
% use \sectionmark{}
% to alter or adjust the section heading in the running head
\vskip15mm
\section{A new theoretical framework  for quantum black holes}
\label{sec:1}

The idea that black holes may explode dates back to Hawking's original paper \cite{Hawking:1974rv}. But Hawking evaporation may not be the primary cause of for black holes to explode. In fact, since then various mechanisms have been proposed that disrupts  the horizon so that matter can be released, possibly in an explosive event \cite{Frolov:1979tu,Hayward:2006}. 
The framework is generic and relies on the possibility that quantum gravity effects would forbid curvature singularity to  develop.  
The works on resolution of cosmological singularity in the context of Loop Quantum Gravity \cite{Ashtekar:2009kx,Rovelli:2013osa} have motivated a recent model for regular black holes \cite{Rovelli:2014kb}. The model imports the main ideas of loop cosmology, in particular that %in the deep quantum regime 
quantum effects can be described at an effective level as a repulsive force. The threshold of the quantum gravitational regime is governed by the energy density rather than by a length, implying that the minimal size that a collapsing object can reach is typically many order of magnitude greater than the Planck length \cite{Rovelli:2014kb}. 

These quantum gravity effects are expected to dominate over Hawking radiation, which can be disregarded in a first order approximation. In such an approximation the equation of General Relativity are invariant under time reversal. 
% Neglect Hawking radiation ?in the first approximation
% Energy conservation at infinity: elastic bounce
% GR is time reversal invariant!
Therefore the black-hole evolution is then described by gluing together a collapsing and an expanding solution of the Einstein equations via a quantum region, where those equations are not satisfied as quantum effect modifies the classical geometry. The process of passing trough a classically forbidden region can be thought as a tunnelling process. In other words, quantum gravity may allow black holes to ``decay'' in a white holes %The association of the late life of a black hole with a white hole 
\cite{Stephens:1993an,Hajicek:2001yd,Haggard:2014rza}.
%\cite{Modesto:2010uh}\cite{Bambi:2013caa,Bambi:2013ufa}
%\cite{Bardeen:2014uaa,Hossenfelder:2009fc,}

% If the width of the figure is less than 7.8 cm use the \texttt{sidecapion} command to flush the caption on the left side of the page. If the figure is positioned at the top of the page, align the sidecaption with the top of the figure -- to achieve this you simply need to use the optional argument \texttt{[t]} with the \texttt{sidecaption} command

%\begin{figure}[h]
%\sidecaption
%%\centerline
%{\includegraphics[height=65mm]{fig/trev2.pdf}}
%\caption{An explicit metric describing a black hole evolving into a white hole has been found in \cite{Rovelli2014-2}. A null shell coming from the past infinity collapses, forms a black hole, bounces, form a white hole and expands toward future infinity.
%The region I is Minkowski spacetime. 
%The region II is locally isomorphic to Schwarzschild.
%The surface separating region II and III is a trapping horizon, and its time-symmetric counterpart is an anti-trapping horizon.
%The region III is a quantum region where Einstein equation are not satisfied.
%}
%\label{ps3}
%\end{figure}

\section{How long does a black hole live?}
For an observer comoving with the collapse, the process is very short: it is just the time light takes to travel in a distance equal to the black holes size. For a solar mass black holes, this is of the order of the milliseconds. On the other hand, for an observer sitting out of the black hole, the process appears redshifted: this redshift, that in the classical theory is infinite, is finite here%
\footnote{The relation between the time inside the horizon and the time outside is coded in the metric. There exist a one-parameter family of metrics modelling the black-to-white process. The extreme case for which the time inside is equal to the time inside has been studied in   \cite{Barcelo:2015uff,Barcelo:2016hgb}.}.
 The value of such a redshift is governed by quantum gravity effects, and can be given in terms of a probability distribution rather than as an exact value. The phenomenological properties of this process depends on this time, the black holes lifetime, that can be expressed as a function of the black-hole total mass. 

The lifetime $\tau$ of the black hole can be constrained by the following heuristic arguments. On the one hand, the ``firewall'' argument \cite{Almheiri:2012rt} provides a time upper bound. This can be see as a no-go theorem involving the following hypothesis: the unitarity of the quantum evolution, the equivalence principle at the base of general relativity and the validity of quantum field theory on a (fixed) curved background. At the Page time (that can be roughly identified with the time after which the mass of the black hole has half evaporated, and is therefore of the order $\sim m^3$  in natural units) the three hypothesis cannot hold together: a signal that the approximation of a fixed background should be abandoned for a fully dynamical theory of the quantum gravitational field. Therefore quantum gravity should manifest, in the form of the decay of the black hole into a white hole, no later than a time $\tau_{max}\sim m^3$.

On the other hand, quantum gravity effects require a minimal time to manifest. In particular, we want the time to be long enough for quantum effect to manifest outside the horizon in order to modify its classical behaviour. The fact that quantum gravity effects can manifest outside of the horizon may sound surprising: we usually identify the quantum gravity regime with a regime of Planckian curvature ($R \sim \ell_{Planck}^{-2}$) but at the horizon the curvature may be small. Consider instead the combination of the curvature and a time:     $R \tau \sim \ell_{Planck}^{-1}$. If now we substitute the value of the curvature near the horizon $R= (2m)^{-2}$ we find that the hole lifetime $\tau$ must be longer or of the order of $\tau_{min} \sim m^2$.

Notice that to determine the black hole lifetime it is required a fully non-perturbative quantum gravity computation. Preliminary results have being obtained in the context of the covariant formalism of Loop Quantum Gravity (spinfoam) \cite{Christodoulou:2016vny} and they seems to indicate that the probability for the black hole decay should be peaked on its shortest permitted value, i.e. $\tau= ~ m^2$, as expected for other known decaying phenomena.

For the following analysis of the phenomenology associated to black-to-white hole decays, we have considered the full window of possible lifetimes $m^2 \lesssim \tau \lesssim m^3$. We parametrize this interval by introducing a parameter $k$ such that $
\tau=4k M^2$.

\section{Primordial black holes and their signature}
\label{sec:2}
The framework of the black-to-white transition is expected to apply to any kind of black hole, irrespectively of the mass or the way it formed. As the lifetime depends on the mass, stellar black hole will explode in the future. To observe an explosion today, we need black holes that are sufficiently small and sufficiently old. PBH satisfy these conditions. The formations of black holes in the early universe can be achieved by a variety of models (see \cite{green} for a review). Here we consider in particular the case of overdense regions collapsing at the time of reheating, but our picture will not qualitatively change for other models of formation, as for instance during the contracting phase in the pre-big-bang epoch \cite{Carr:2011hv}, as the long cosmological times dominate over short differences in the exact time of formation.

\begin{figure}[h!]
\sidecaption
\includegraphics[height=40mm]{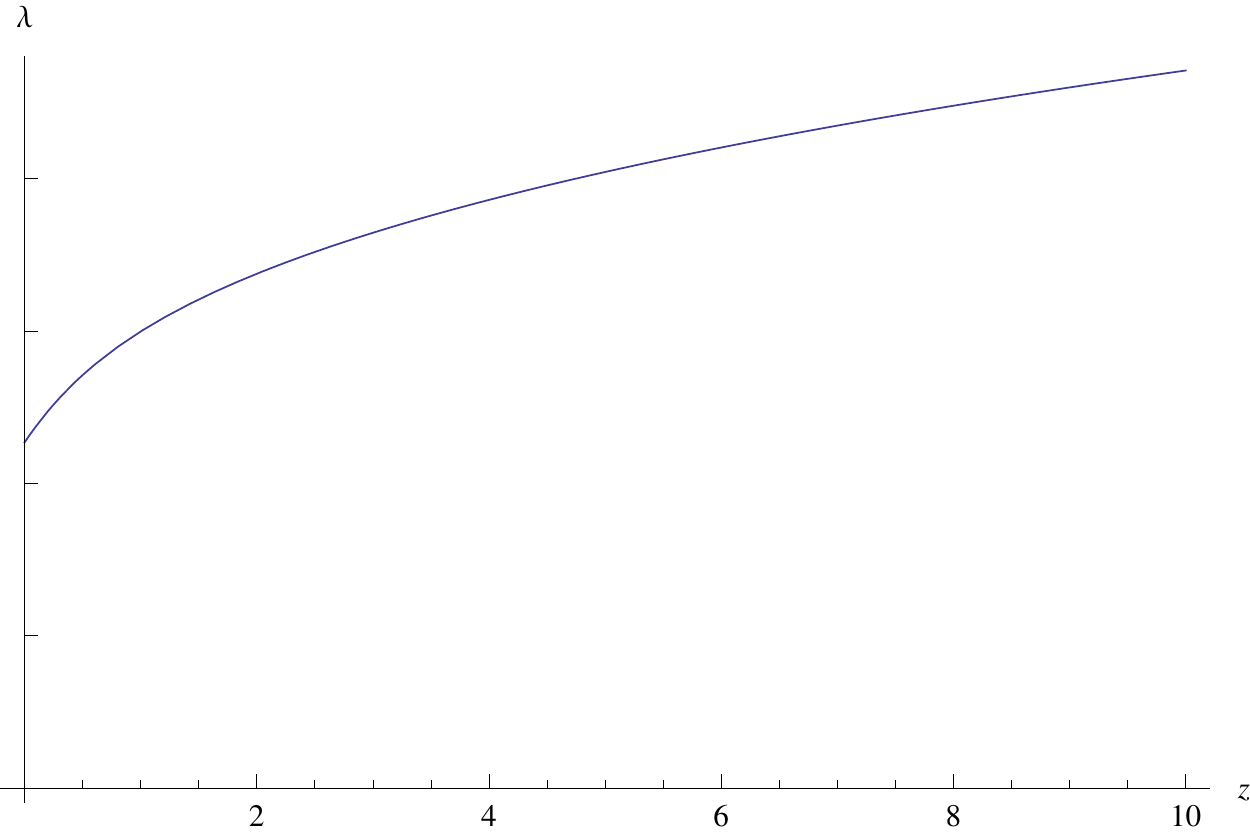}
\caption{The expected wavelength (unspecified units) of the signal from black hole explosions as a function of the redshift $z$. The curve flattens at large distance: the shorter wavelength from smaller black holes exploding earlier get compensated by the redshift. %This provide a quantum gravity signature.
}
\label{fig:flat}
\end{figure}

But if PBH explode, how can they be distinguished from all the other astrophysical sources? Remarkably, their signal carries a characteristic signature. The wavelength of the signal depends on the mass of the exploding black hole%
\footnote{Notice that this differs from the case of black holes exploding via Hawking evaporation, as in that case they would all explodes when they reach the Planck size irrespectively of their initial mass. }.
Smaller PBH should have exploded earlier: smaller black holes produce a signal of shorter wavelength, but this get redshifted as we can observe them as an earlier (and therefore distant) explosion. It is possible to compute how the wavelength scales with the distance (in terms of the redshift). Standard astrophysical objects scale with a simple linear law. Instead, we find a peculiar flat curve \cite{Barrau2014-2}  where the shorter wavelengths get compensated by the amount of redshirt (Fig.\ref{fig:flat}). Ideally, we would like to detect very energetic burst for which the distance of the source can be known, in order to fit this curve.

% distant signals ?originated in younger ?and smaller sources
% distance ? 1/wave length
% taking into account the redshift the resulting function is very slowly varying

\subsection{Description of the expected signals}

The model of black-to-white tunnelling provides a concrete mechanism for the explosion, but lacks of any detail of the precise astrophysical process. Heuristic arguments lead us to consider two possible signal channels, that may concur together to the total emission. Both depend on the initial mass of the black hole, but for different reasons.

The first one, that we call the \emph{high energy} channel, is given by matter (mainly photons) that is re-emitted at the same temperature it had at the time it collapsed forming the black hole, as its co-moving bouncing time is very short. In the simplest model, PBH form at different sizes corresponding to the Hubble horizon as the universe expands and cools. This happens typically at the reheating, therefore at a temperature of the order of $TeV$. This suggests a high energy component of the signal in the order of $TeV$.   This is a very interesting observational window, as $TeV$ astronomy is expected to develop in the forthcoming year \cite{Rieger:2013awa}. On the other hand, our observational horizon is limited: cosmic rays at such an high energy interact with the cosmic background radiation. Therefore the high energy channel would be observable only for events happening in our galaxy or nearby.

The second one, denoted as \emph{low energy} channel, assumes that the signal will carry a mode that corresponds to the size (i.e. the Schwarzschild radius, that is just $R=2m$) of the exploding object. Knowing the PBH lifetime $\tau$, we can estimate the mass of black holes exploding today. %\\
For $\tau\approx m^3$, the emitted signal would be in the $GeV$, but a detailed analysis \cite{Barrau2014} of such a signal have shown that $MeV$ photons will have higher density and are more likely to be detected. Transient signals in this range are Short Gamma Ray Bursts \cite{nakar}, whose origin is still unclear.
As their energy is so high, the dispersion due to the cosmic background limits our observational horizon .%(see Fig.\ref{fig:distance}).
%%
%\begin{figure}[b]
%    \centering
%        \hspace{-4mm}\includegraphics[width=.5\textwidth]{fig/Distance_LowEnergy.pdf}
%    ~~ %add desired spacing between images, e. g. ~, \quad, \qquad, \hfill etc. 
%      %(or a blank line to force the subfigure onto a new line)
%        \includegraphics[width=.5\textwidth]{fig/Distance_HighEnergy.pdf}
%\caption{Maximum distance at which a bouncing black hole can be detected in the {\it low energy} (left) and {\it high energy} (right) channel as a function of the parameter $k$ (and the associated signal energy).}
%    \label{fig:distance}
%\end{figure}
%%
For $\tau\approx m^2$, the estimated signal is expected in the millimetre range of the radio spectrum. Interesting, 
very energetic transients (Fast Radio Bursts \cite{Katz:2016dti}) have recently being discovered in the radio frequency; these   may candidate as detection of PBH explosions. Given the approximations taken in the present model, the energy of the Fast Radio Bursts are intriguingly close to those predicted by this model. The radio window allows for observational depth, giving virtually access to events in the entire observable universe. Large antenna available on earth would detect even faint signals in the longer radio wavelength. On the other hand, the (sub)millimetre  wavelengths are shielded by the atmosphere: for them we relay on space telescope, whose detection technology is not sensitive to transients. 
A more promising strategy seems to be to study the integrated emission, i.e. the relic radiation from all the detectable past explosions.

\newpage
\subsection{Diffuse Emission}
\label{subsec:3}

\begin{figure}[t]
    \centering
        \hspace{-2mm}
        \includegraphics[width=.47\textwidth]{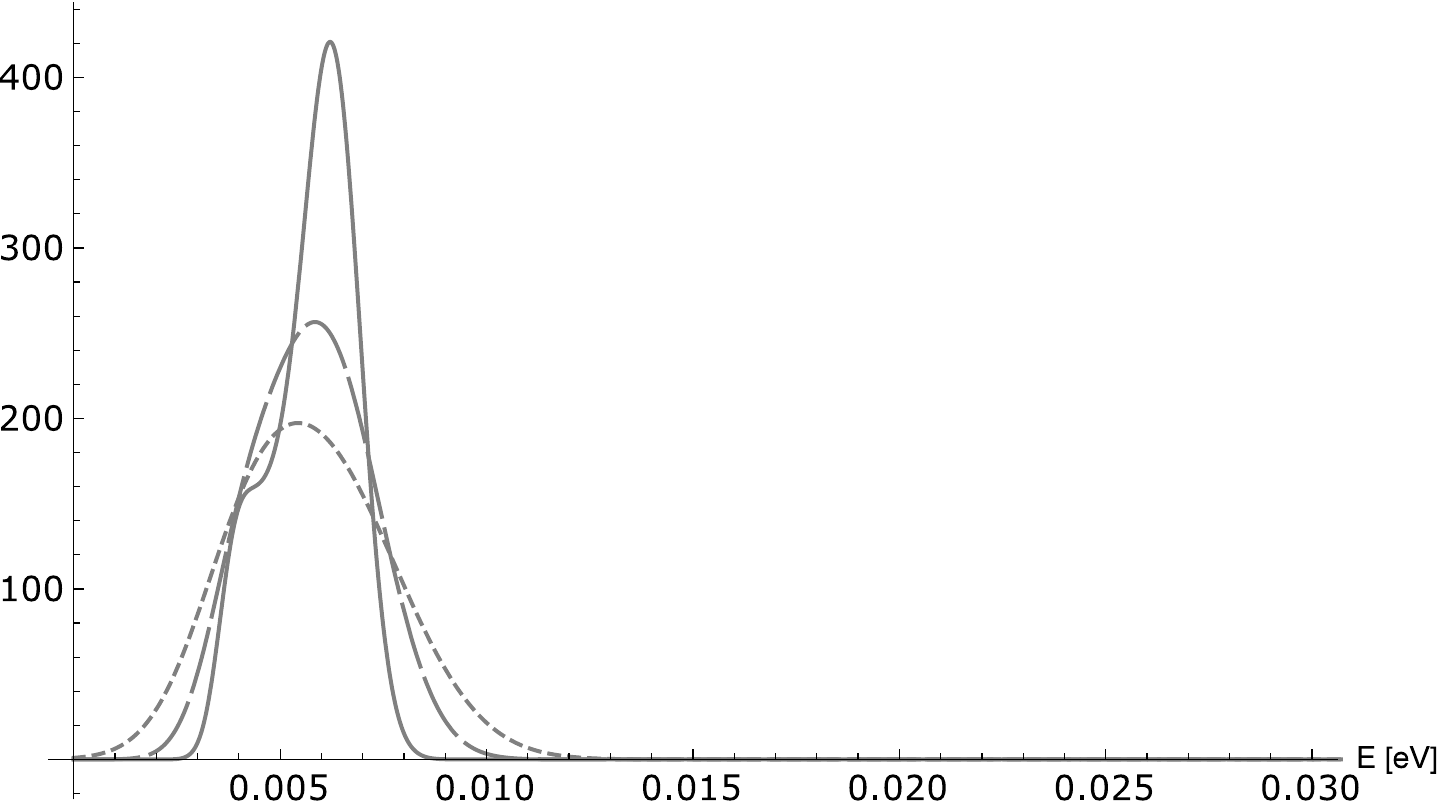}
    \hskip5mm %add desired spacing between images, e. g. ~, \quad, \qquad, \hfill etc. 
        %\includegraphics[width=.48\textwidth]{fig/high05.pdf}
%\caption{The diffuse emission of the low energy (left) and the high energy (right) emission plotted for the minimal $k$, i.e. for the shortest lifetime.}
%    \label{fig:diffuse05}
%\end{figure}
%\vspace{-1cm}
%\begin{figure}[h]
%    \centering
%        \hspace{-1mm}
        \includegraphics[width=.47\textwidth]{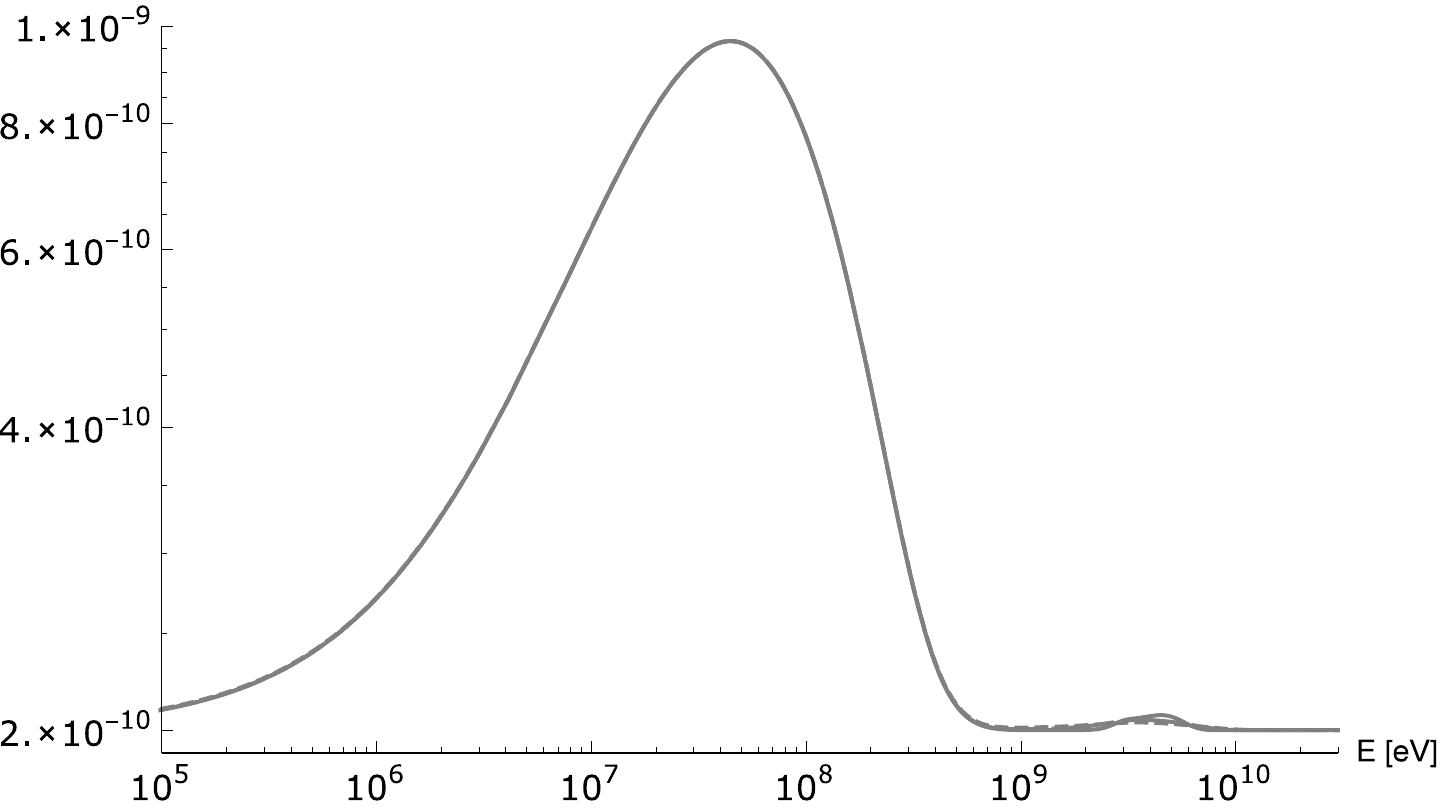}
%    ~~~ %add desired spacing between images, e. g. ~, \quad, \qquad, \hfill etc. 
%        \includegraphics[width=.48\textwidth]{fig/high22.pdf}
%\caption{The diffuse emission of the low energy (left) and the high energy (right) emission plotted for the maximal $k$, i.e. for the longest lifetime.}
\caption{The diffuse emission of the low energy channel plotted for the minimal $k$, i.e. for the shortest lifetime (left), and for the maximal $k$, i.e. for the longest lifetime (right). The plots have being obtained using the PYTHIA code \cite{tor}, that gives the particle production for a process at a given initial energy.}
    \label{fig:diffuse}
    %\vspace{-2em}
\end{figure}

A detailed study of the diffuse emission from PBH has been carried in \cite{Barrau:2015uca} considering the full range of possible PBH lifetime, for both the proposed channels of emission. The resulting spectrum carries a distortion from the expected black-body spectrum due to the characteristic redshift-wavelength relation of Fig.\ref{fig:flat}. We report here (Fig.\ref{fig:diffuse}) a sample of our results. Units in the ordinate axis are not better specified as the normalization of the spectrum depends on the amount of PBH contributing to dark matter.
%\vspace{-1em}
\begin{figure}[b]
    \centering
\sidecaption
%\vspace{-6em}
\includegraphics[width=0.7\textwidth]{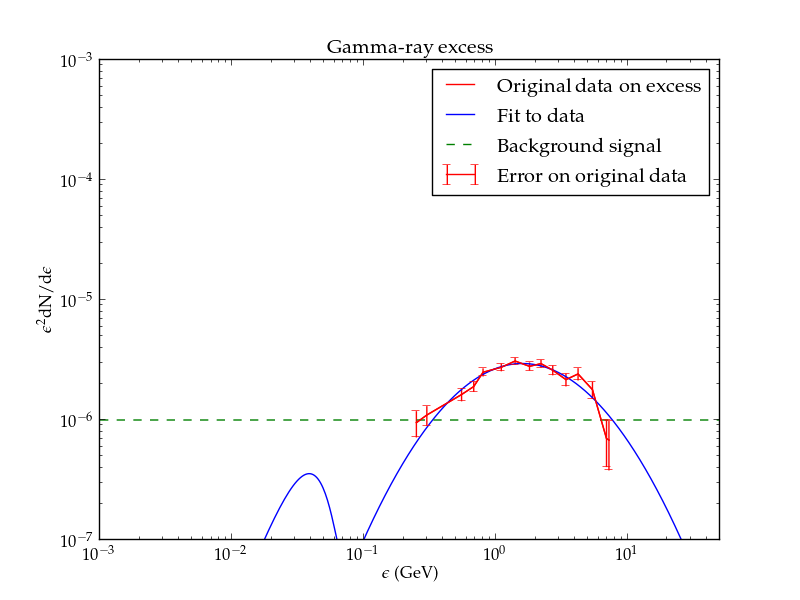}
\caption{Spectral energy density $(\epsilon^2dN/d\epsilon)$ fiting the Fermi-LAT data by exploding PBH. %The fit is obtained for a PBH lifetime value closer to the maximal allowed. 
The agreement with data is good, with a $\chi^2$ per degree of freedom of $1.05$. The bump on the left is given by the secondary gamma-rays, whose spectral energy density is much lower than the one of primary photons and remains below the background (dash line).}
\label{fig:fit}
\end{figure}

The study of the diffuse emission provides a promising tool to constrain the model and may lead to some unexpected surprises. One may ask whether the measured value of the background radiation can be explained by considering a contribution from exploding PBH. We have tested this possibility for the excess of gamma rays coming from the galactic center measured by the Large Area Telescope (LAT) installed in the  {\it Fermi} satellite \cite{Daylan:2014rsa}. We found \cite{Barrau:2016fcg} a remarkable good fit for the case of PBH with a lifetime close to the maximal allowed by the model (Fig.\ref{fig:fit}). 

\pagebreak
The phenomenology of PBH exploding via a quantum gravity black-to-white transition has just started to be explored. It presents peculiar new features that have consequences for PBH dark matter models. They differs from the ones given by explosions via Hawking evaporations and they provide a new windows for quantum gravity phenomenology. 

\vspace{2em}
%We stress that the explanation we suggest is specifically associated with the quantum gravity scenario considered in this work. The time integrated spectrum of black holes evaporating by the Hawking process is scaling as $E^{-3}$ and there is no way it can explain the {\it Fermi} excess. As explained before, two  parameters are required to fully determine the low-energy component of bouncing black holes: their bouncing time and the width of the primary Gaussian.  

%%
%\begin{acknowledgement}

%\end{acknowledgement}
%%

% BibTeX users please use
% \bibliographystyle{}

\bibliography{bib}

\end{document}